\documentclass{elsart}
\usepackage{natbib}
\usepackage{graphicx}

\usepackage{amssymb}
\usepackage{amsmath}
\usepackage{bm}

\begin{document}

\begin{frontmatter}
\title{Baryons and weak lensing power spectra}
\author[address1]{Martin White\thanksref{mwemail}},
\address[address1]{Departments of Physics and Astronomy,
University of California\\Berkeley, CA, 94720}
\thanks[mwemail]{E-mail: mwhite@berkeley.edu}

\begin{abstract}
We provide a simple toy model to elucidate the effects of
baryonic cooling on predictions of the convergence power spectrum
in weak gravitational lensing.
Our model suggests that the effects of baryonic cooling on
dark matter halo profiles can alter the dark matter power
spectrum at $k\sim 10\,h\,{\rm Mpc}^{-1}$ and lead to percent
level changes in the convergence angular power spectrum at
wavenumbers of several thousand.
These effects would be measurable with planned weak lensing
surveys and will impact constraints on cosmological parameters.
\end{abstract}

\end{frontmatter}

\section{Introduction} \label{sec:intro}

The study of modern cosmology has been tremendously advanced by
probes for which detailed comparison of theory and observation
is possible.  The prototypical example is the anisotropies in
the cosmic microwave background (CMB) for which theoretical
calculations have reached a highly refined level \citep{SSWZ}
and measurements a high degree of accuracy \citep[e.g.][]{WMAP}.
Weak gravitational lensing, the small shearing of the shapes of
background galaxies by gravitational lensing from large-scale
structure along the line-of-sight, is another.
With detections of the effect now becoming possible and a relatively
clean theory \citep{LensReview1,LensReview2}, gravitational lensing
is beginning to become a critical cosmological probe, already providing
constraints on the mass density and the fluctuation amplitude
\citep[see e.g.][for the current status]{WaeMel,HoeYeeGla} 
and touted for its potential to constrain other interesting cosmological
parameters.

One of the reasons often quoted for the cleanness of weak gravitational
lensing is that the deflection of light involves only gravity, and this
circumvents many of the intractable problems associated with complex
baryonic physics (e.g.~cooling, feedback, star formation etc.).
Clearly this statement is an idealization.  The ability of baryons to
cool and collapse into dense structures alters the gravitational field
around them and will thus alter the predictions of gravitational lensing.
The dense cores of galaxies can exceed the critical density necessary
for the production of multiple images, strongly modifying the tail of
the lensing distribution.  The cooling of baryons also causes a
concentration of the halos in which they live, altering the matter power
spectrum on small scales.

We wish to investigate this latter effect in this paper, using an
analytic toy-model to gain a feel for the expected amplitude.  This
will allow us to estimate the ``theory uncertainty'' associated with
the predictions, and provide guidance on experimental design.
As we shall see, precise measurements of the small-scale lensing power
spectrum will be sensitive to the details of baryonic physics, providing
another interesting probe of these complex phenomena.

\section{Relevant scales} \label{sec:scales}

We begin by discussing the relevant length and angular scales before
turning to our toy model for including baryonic cooling.
We shall work throughout in terms of the power spectrum of the convergence,
$\kappa$.  This, or equivalent measures, has received much of the
attention from the community to date.  The convergence is a weighted
integral of the mass density along the line-of-sight:
\begin{equation}
  \kappa \simeq \frac{3}{2}H_0^2\Omega_{\rm mat} \int d\chi
  \ g(\chi) \frac{\delta}{a}
\end{equation}
where $\delta$ is the overdensity, $a$ is the scale-factor, $\chi$ is the
comoving distance and $g(\chi)$ is the lensing weight
\begin{equation}
  g(\chi) \equiv \int_\chi^\infty d\chi_s p(\chi_s)
    \frac{\chi(\chi_s-\chi)}{\chi_s}
\label{eqn:g(chi)}
\end{equation}
for sources with distribution $p(\chi_s)$ normalized to $\int dp=1$.
Using the Limber and Born approximations we can write the (dimensionless)
power spectrum of $\kappa$ as an integral over the mass power spectrum
\citep{LensReview1,LensReview2}
\begin{equation}
  \Delta_{\kappa}^2 (\ell) = \frac{9\pi}{4\ell} \Omega_{\rm mat}^2 H_0^4
  \int\,\chi'\,d\chi'\
  \left[\frac{g(\chi')}{a(\chi')}\right]^2 \Delta_{\rm m}^2(k=\ell/\chi,a)
\label{eqn:massspectrum}
\end{equation}
where $g(\chi')$ is defined in Equation \ref{eqn:g(chi)},
$\Delta_{\rm m}^2(k) = k^3 P(k)/(2 \pi ^2)$ is the contribution to the
mass variance per logarithmic interval in wavenumber and
$\Delta_{\kappa}^2 (\ell) = \ell^2 C_\ell /(2 \pi)$ is the contribution to
$\kappa_{\rm rms}^2$ per logarithmic interval in angular wavenumber $\ell$.

Our ability to measure $\Delta^2_\kappa$ is constrained on both large
and small scales.  On large scales the finite amount of sky we can cover
leads to a ``sample variance'' error.  For a Gaussian $\kappa$ field,
a very good approximation on scales above a degree, this leads to a
fractional error
\begin{equation}
  \frac{\delta C_\ell}{C_\ell} = \sqrt{\frac{2}{(2\ell+1)f_{\rm sky}}}
\end{equation}
where $f_{\rm sky}$ is the fraction of sky surveyed.  On small scales
we are limited by the finite number of galaxies with which we can
measure the shear.  If $\bar{n}_{\rm gal}$ is the number density of
galaxies of intrinsic ellipticity $\gamma_{\rm rms}$, this shot-noise
term increases the error to \citep{LensReview1,LensReview2}
\begin{equation}
  \frac{\delta C_\ell}{C_\ell} = \sqrt{\frac{2}{(2\ell+1)f_{\rm sky}}}
  \left( 1 + \frac{\gamma^2_{\rm rms}}{\bar{n}_{\rm gal}C_\ell} \right)
\end{equation}
Ambitious experiments now being planned hope to measure $\Delta^2_\kappa$
over the range $10^2\le\ell\le10^4$ using many tens of galaxies per square
arcminute over much of the sky\footnote{see e.g.~http://pan-starrs.org/ ,
http://www.ctio.noao.edu/telescopes/dec.html ,
http://www.lsst.org and http://snap.lbl.gov}.
For an all-sky experiment with $\bar{n}_{\rm gal}\to\infty$ we could
measure the power in a 10\% band around $\ell=10^3$ to $0.3\%$.
This motivates us to consider what the theoretical uncertainty on our
predictions is.

\begin{figure}
\begin{center}
\resizebox{5in}{!}{\includegraphics{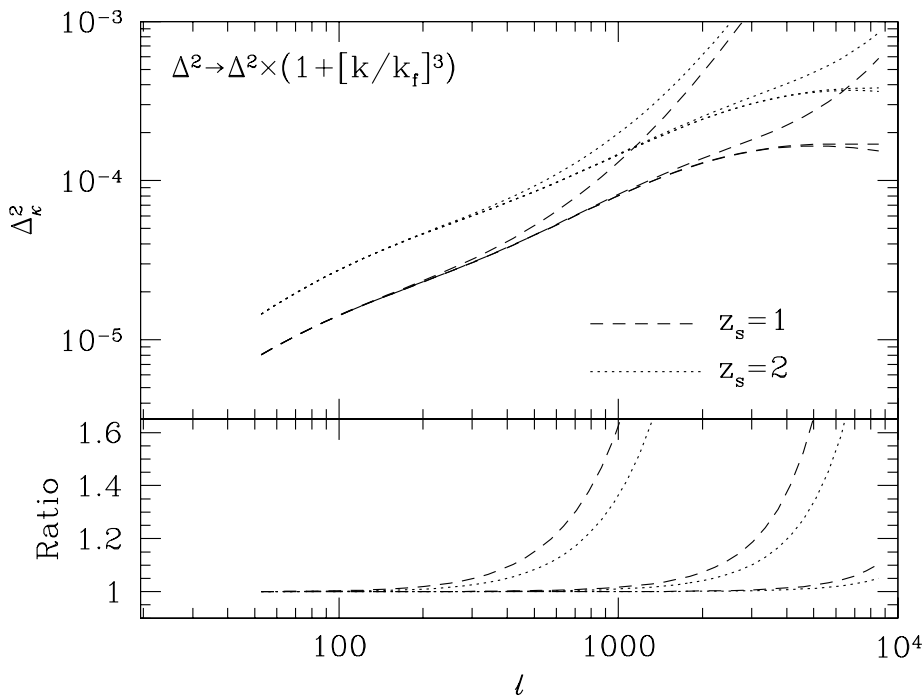}}
\end{center}
\caption{The effect of (artificially) increasing the small scale mass
power spectrum on the angular spectrum of $\kappa$.  The top panel shows
the angular power spectra for sources at $z_s=1$ (dashed) and 2 (dotted)
when $\Delta_{\rm mass}^2(k)$ is multiplied by $1+(k/k_f)^3$ for
$k_f=\infty$, 30, 10 and $3\,h\,{\rm Mpc}^{-1}$ (in order of increasing
small scale power).  The lower panel shows the power ratio, compared
to $k_f=\infty$.}
\label{fig:filter}
\end{figure}

\begin{figure}
\begin{center}
\resizebox{5in}{!}{\includegraphics{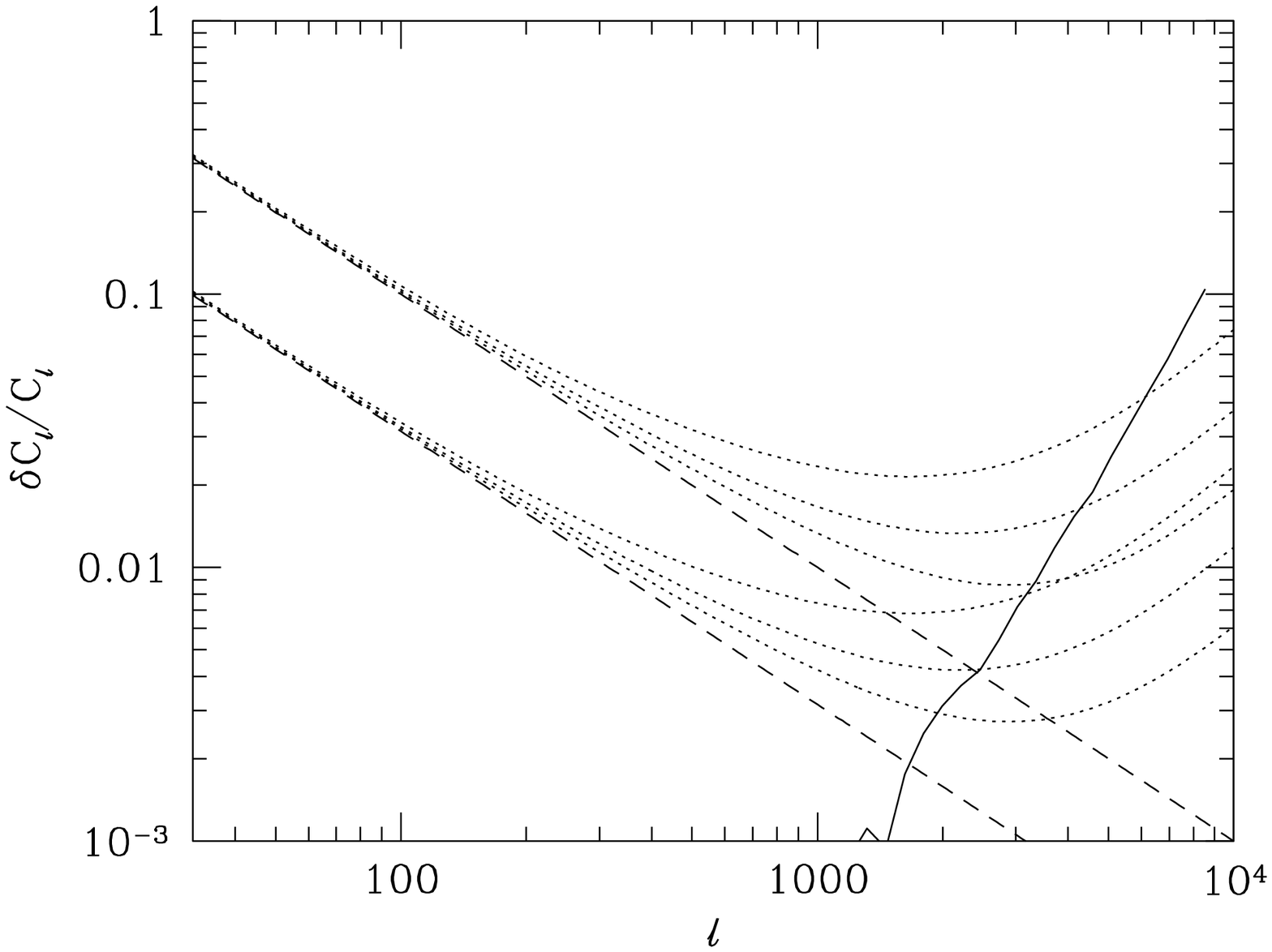}}
\end{center}
\caption{The fractional error in $C_\ell$, vs.~multipole $\ell$ for
sources at $z_s=1$.  The solid line shows the shift in $C_\ell$ for
$k_f=30\,h\,{\rm Mpc}^{-1}$ from Figure \protect\ref{fig:filter}.
The results of section \ref{sec:halo} suggest this is a good ballpark
estimate for the effect of baryons.
The dashed lines show the cosmic variance limit for Gaussian fluctuations
in a band of width $\Delta\ell=0.1\,\ell$ for $f_{\rm sky}=0.1$ (upper)
and $1$ (lower).
The dotted lines include shot noise for galaxy number densities
$\bar{n}_{\rm gal}=100$, 50 and 25 galaxies per square arcminute
(bottom to top) assuming the fiducial model, $k_f\to\infty$.}
\label{fig:dcl}
\end{figure}

\section{A model}

Previous work \citep[e.g.][]{JSW,ValWhi,WhiVal} has studied the level
of accuracy of assumptions in the Limber and Born approximations and
numerical artifacts arising from the calculational procedure under the
assumption that only gravitational physics operated.  Here we want to
consider the effects of baryonic physics, specifically cooling, on the
power spectrum.
To set the stage we show in Figure \ref{fig:filter} how a (redshift
independent) change in the matter power spectrum translates into a
change in the angular power spectrum of $\kappa$ through
Equation \ref{eqn:massspectrum}.
We have artificially increased the small scale power by multiplying
$\Delta_{\rm mass}^2$ by $1+(k/k_f)^3$ for a range of $k_f$.
Note that setting $k_f=30\,h\,{\rm Mpc}^{-1}$ corresponds to a 4\%
increase in power at $k\simeq 10\,h\,{\rm Mpc}^{-1}$, while setting
$k_f=10\,h\,{\rm Mpc}^{-1}$ is a factor of 2 increase on the same scale.
The comparison of the shift in the signal to the estimated uncertainty
is given in Figure \ref{fig:dcl} for sources at $z_s=1$.

At what scale do we expect baryonic physics to be important?  The
dominant effect of hot baryons is to isotropize the inner regions of
halos, due to the finite (isotropic) pressure they experience.  This
has a negligible effect on the power spectrum.  Thus we are more
interested here in the baryons which can cool.  
We shall assume that the dominant effect of cooling is to cause the
baryons to contract, potentially dragging some of the dark matter with
them through their gravitational coupling.
Since substructures make up a small fraction of the total mass of a halo
we shall consider only the baryons cooling to the center of any given
halo, which should dominate the effect.

\begin{figure}
\begin{center}
\resizebox{5in}{!}{\includegraphics{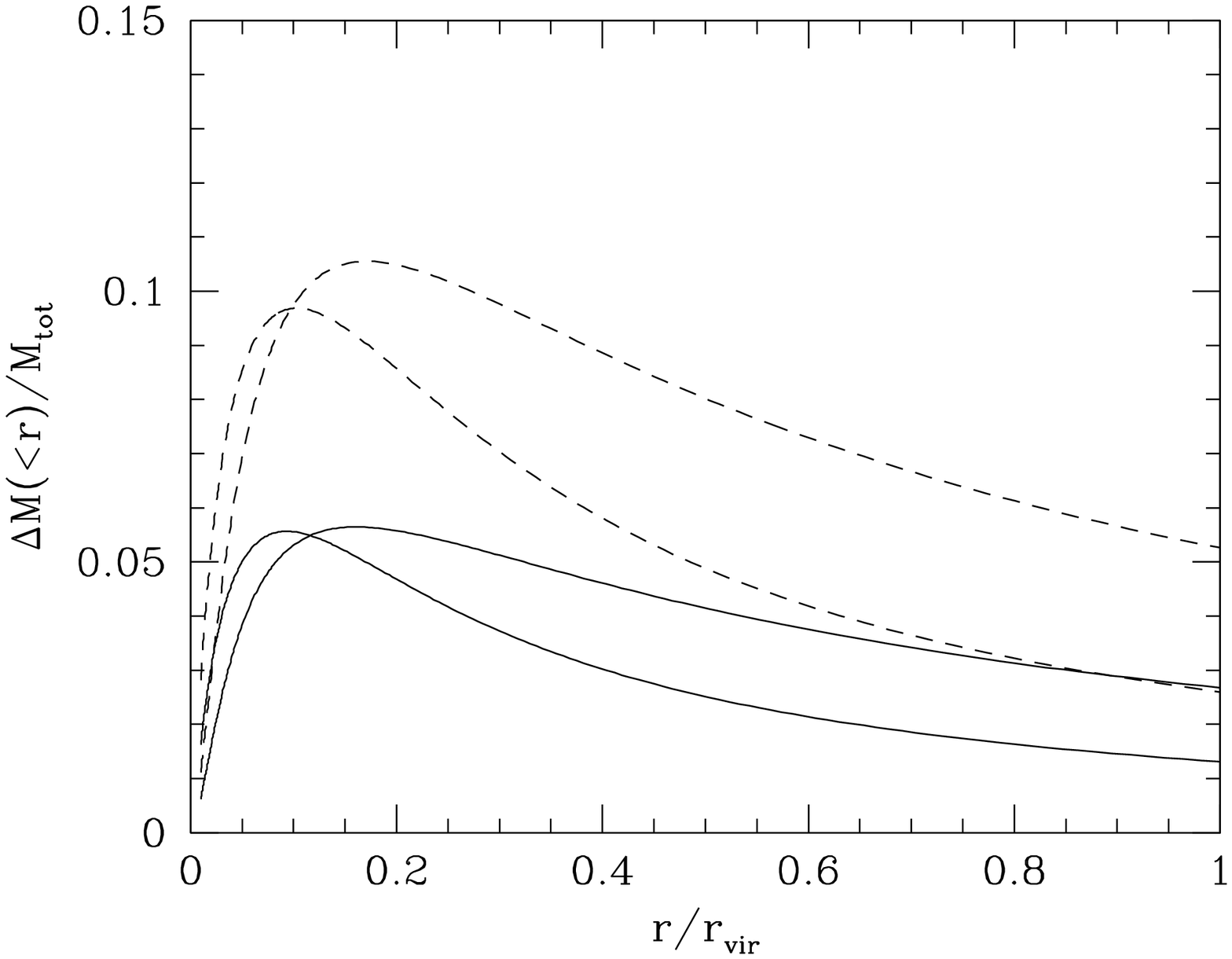}}
\end{center}
\caption{The effect of adiabatic cooling on NFW halos.  The change in the
enclosed mass is plotted versus radius.  The solid lines are for
$f_{\rm cool}=5\%$ while the dashed lines are for $f_{\rm cool}=10\%$.
The lower curves in each case have $c=5$, $a_i=0.1\,r_{\rm vir}$ and
$a_f=0.1\,a_i$ while the upper curves have $c=10$ and $a_i=0.3\,r_{\rm vir}$.}
\label{fig:adiabatic}
\end{figure}

To gain a feel for the size of the effect of cooling on the mass profiles
of halos we make use of the adiabatic cooling model.
We assume ``sphericized'' halos containing dark matter and hot and cold
baryons.  We ignore the pressure of the hot baryons, treating them in the
same manner as the (cold) dark matter.  The initial halo is modeled
as an NFW profile \citep{NFW} with
\begin{equation}
  M(<r) \propto \log(1+x)-\frac{x}{1+x}
\end{equation}
where $x=cr/r_{\rm vir}$ is a scaled radius for a halo of concentration $c$.
The baryons are assumed to follow a Hernquist profile \citep{Hernquist}
\begin{equation}
  M(<r) \propto \frac{x^2}{(1+x)^2}
\end{equation}
with $x=r/a$.  Half of the mass is contained within $r\simeq 2a$ in this model.

For a massive halo, say a rich cluster of galaxies, the concentration is near
$5$ and the virial radius is $\sim 1.5\,h^{-1}$Mpc while the cooling radius is
$O(100\,{\rm kpc})$ so we take the initial scale $a_i=0.05-0.1\,r_{\rm vir}$.
Cooling usually allows the baryons to contract by a factor of 10 in linear
scale, so we take $a_f=0.1\,a_i$.
Our final results are quite insensitive to the choice of $a_i$ within
this range, and even to the choice of the other parameters to a large
extent.

Adiabatic contraction conserves both the mass and the angular momentum of
the cold baryons and dark matter particles \citep{BFFP,KocWhi}.
Thus given a final radius we can solve for the initial radius through 
\begin{eqnarray}
  M_f(r_f)     &=& M_i(r_i) + M_{cf}(r_f) - M_{ci}(r_i) \\
  r_f M_f(r_f) &=& r_i M_i(r_i)
\end{eqnarray}
where $M_i$ and $M_f$ are the initial and final total mass profiles and
$M_{ci}$ and $M_{cf}$ are the initial and final distributions of the
material which can cool.
We show the effects of this in Figure \ref{fig:adiabatic}.
The cooling and contraction of the baryons draws material into the center,
enhancing the central density.  The effect is not very sensitive to
the initial halo concentration and scales almost linearly with the cooled
mass fraction.

\section{Effect on mass power spectrum} \label{sec:halo}

To see what the effect of these changes are on the matter power spectrum
we use the halo model \citep[see][for a review]{CooShe}.
While this will not serve for high precision cosmology, it will illustrate
the typical scale of the effects.
The fundamental assumption in this model is that all of the mass in the
universe resides in virialized halos whose clustering properties are those of
peaks of the density field and whose profiles are drawn from a universal one
parameter family depending only on halo mass.
The power spectrum is then made up of two contributions
\begin{equation}
  P(k) = P^{{\rm 2-halo}}(k) + P^{{\rm 1-halo}}(k)
\end{equation}
one from pairs of mass elements which lie in different halos and one where
they both lie in the same halo.
The first term dominates on large scales, while the second dominates on
the smaller scales of interest to us.
Within this model we can compute the mass weighted average of any quantity
by integrating over the multiplicity function
\begin{equation}
  f(\nu) d\nu = \frac{M}{\bar{\rho}} \frac{dn}{dM} dM
\end{equation}
where $\nu$ is the peak height.
For $f(\nu)$ we use the fit to N-body simulations quoted in \citep{SheTor}.

The power spectrum from the 2-halo term is that due to a system of (smooth)
halos of profile $y(k)$ laid down with inter-halo correlations assumed to
be a biased sampling of $P_{\rm lin}(k)$.
This dominates on large scales.
Since the real space convolution is simply a Fourier space multiplication
this contribution is
\begin{equation}
  P^{{\rm 2-halo}}(k) = P_{\rm lin}(k) \left[
  \int f(\nu) d\nu\ b(\nu) y(k; M) \right]^2
\label{eqn:twohalo}
\end{equation}
where $b(\nu)$ is the (linear) bias of a halo of mass $M(\nu)$ which we take
{}from \citep{SheTor}.
We assume that the halos all have spherical profiles depending only on the
mass.  We neglect any substructure or halos-within-halos.
We take the NFW form with Fourier transform
\begin{equation}
\begin{array}{ll}
  \widetilde{\rho}(k) = 4\pi\rho_0 r_s^3 \left[ \vphantom{\int} \right. &
    \cos z \left\{ {\rm Ci}([1+c]z) - {\rm Ci}(z) \right\} +  \\
&  \sin z \left\{ {\rm Si}([1+c]z) - {\rm Si}(z) \right\} -
   \left. { \displaystyle{ \frac{\sin cz}{(1+c)z}} } \right]
\end{array}
\end{equation}
where $z\equiv kr_s$ and Ci and Si are the cosine and sine integrals.

On small scales we are dominated by pairs lying within a single halo
\begin{equation}
  P^{{\rm 1-halo}}(k) = \frac{1}{(2\pi)^3} \int f(\nu) d\nu
  \ \frac{M(\nu)}{\bar{\rho}} |y(k)|^2
\label{eqn:onehalo}
\end{equation}
where $\bar{\rho}$ is the mean matter density.  This term gives the major
contribution to the angular power spectrum of $\kappa$ on the scales of
interest.

\begin{figure}
\begin{center}
\resizebox{5in}{!}{\includegraphics{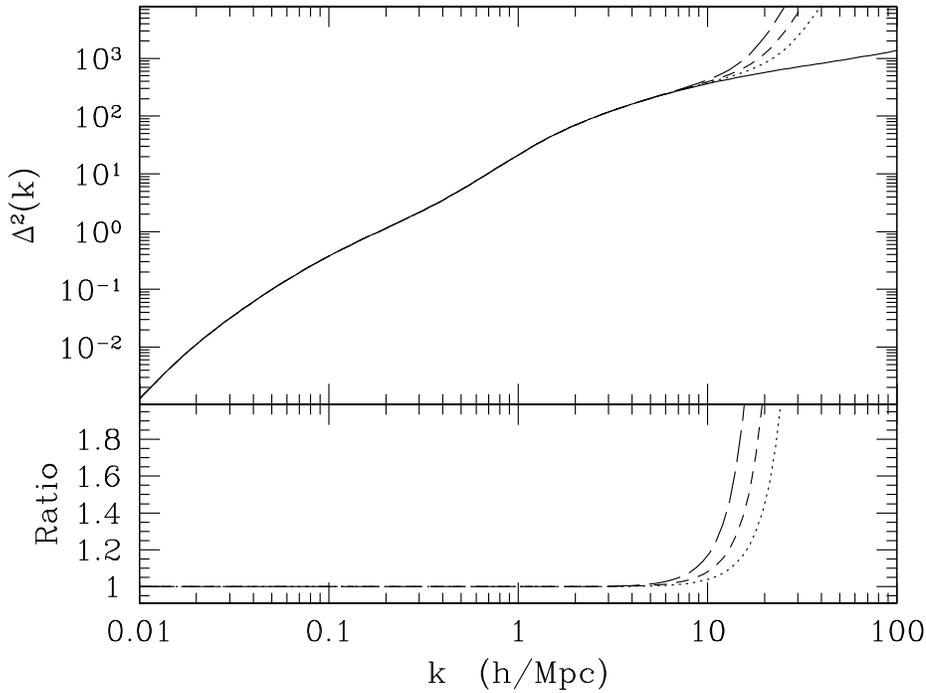}}
\end{center}
\caption{The effect of cooling baryons on the mass power spectrum.
The solid line is the unmodified power spectrum, the dotted line includes
the correction for $\epsilon=0.05\tanh(10^{14}\,h^{-1}M_\odot/M_{\rm vir})$
and $r_0=0.05\,r_{\rm vir}$, the dashed line for
$\epsilon=0.1\tanh(10^{14}\,h^{-1}M_\odot/M_{\rm vir})$
and the long-dashed line for
$\epsilon=0.2\tanh(10^{14}\,h^{-1}M_\odot/M_{\rm vir})$ to illustrate the
sensitivity to our assumptions.  The lower panel shows the ratio to the
unperturbed result.  By $z\simeq 1$ the increase is shifted to the right
by a factor $\sim 1.2$.}
\label{fig:dpk}
\end{figure}

Within this formalism it is easy to see how we incorporate the effects
of cooling baryons.  The 1-halo term, which dominates the scales of
interest, is simply modified by $y(k)\to y(k)+\delta y(k)$ where
\begin{equation}
  \delta y(k) = k\int_0^\infty dr\ \frac{\Delta M(<r)}{M_{\rm tot}} j_1(kr)
\end{equation}
for changes in the cumulative mass, $\Delta M(<r)$, which vanish at $r=0$
and infinity.  Here $j_1(x)$ is the spherical Bessel function of order 1.

In keeping with our attempts to construct an analytic model of the effect
we fit $\Delta M(<r)$ with a simple form
\begin{equation}
  \frac{\Delta M(<r)}{M_{\rm tot}} = 
  \varepsilon \left(\frac{rr_0}{r_0^2+r^2}\right)
\end{equation}
which is a reasonable fit to the curves of Figure \ref{fig:adiabatic}.
Then $\delta y(k)$ is
\begin{eqnarray}
  \delta y(k) &=& \varepsilon \frac{\pi}{2}
  (kr_0)\left[ 1-\left(1+kr_0\right)e^{-kr_0} \right]\\
  &=& \varepsilon (\pi/4)(kr_0)^3 + \cdots\qquad kr_0\ll 1
\end{eqnarray}
allowing us to compute $\delta P(k)$ as a function of $\epsilon$ and $r_0$.
It is also possible to find an analytic transform for
$(r/r_0)^3/(1+(r/r_0)^4)$, which has a flatter core as $r\to 0$.  The
large-$r$ and low-$k$ behavior is the same and for the $k$-range of
interest the results are almost indistinguishable.

Fitting the massive halo differences shown in Figure \ref{fig:adiabatic}
we find $\varepsilon\simeq 0.15$ and $r_0\simeq 0.05\,r_{\rm vir}$.
We expect the cooled fraction to be very small in high mass halos and to
increase with mass, until it reaches a limit set by
$\Omega_{\rm bar}/\Omega_{\rm mat}$.
Taking $\varepsilon\sim 3\,f_{\rm cool}$ we model this as
$\varepsilon=(0.03-0.1)\,\tanh(10^{14}\,h^{-1}M_\odot/M_{\rm vir})$.
We also expect $r_0$ to become a larger fraction of $r_{\rm vir}$ for lower
mass halos -- the trend is weak however, so we hold $r_0=0.05\,r_{\rm vir}$
for simplicity.
The precise details of the low mass behavior are unimportant for us, as it
is the most massive halos that dominate the
$k\simeq 10\,h\,{\rm Mpc}^{-1}$ regime which will be our focus.
Also for $kr_0\ll 1$ there is a degeneracy between the effects of a change
in $r_0$ and a change in $\varepsilon$.

Using the $\varepsilon$ and $r_0$ above gives the power spectrum shown
in Figure \ref{fig:dpk}.  In this model there is a few percent increase
in power at $k\simeq 10\,h\,{\rm Mpc}^{-1}$ and a factor of two increase
at $k\simeq 20\,h\,{\rm Mpc}^{-1}$.
By $z\simeq 1$ the curves have shifted to the right by a factor $1.2$.
The ratio is reasonably well fit by the $1+(k/k_f)^3$ form used earlier
with $k_f\simeq 30\,h\,{\rm Mpc}^{-1}$.
Thus Figure \ref{fig:filter} suggests this would lead to a few percent
increase in the power at $\ell\simeq 3000-5000$, which is potentially
measurable in upcoming experiments (see Figure \ref{fig:dcl}).
There is at least a factor of two uncertainty in our estimate of this scale,
but it shows that we expect percent level modifications to the matter
power spectrum at $k\sim 10\,h\,{\rm Mpc}^{-1}$ and the convergence power
spectrum at $\ell\simeq 3000-5000$ from baryonic cooling.

Currently we are unable to reliably compute the effects of cooling, star
formation and feedback on the small-scale mass distribution.
This uncertainty will need to be folded into our requirements on theoretical
calculations of the spectrum, or mitigated in some way.  One possible route
to reducing the sensitivity to the uncertain high-$k$ power is to make use
of multiple source redshifts.  Within the Limber approximation of
Equation \ref{eqn:massspectrum} only modes transverse to the line-of-sight
contribute.  This means that at fixed wavenumber, $\ell$, higher redshift
corresponds to larger wavelength.  By combining signals from a number of
different source redshifts the sensitivity to low-$z$ structure can be
reduced (``nulling tomography''), and with it the sensitivity to high-$k$
at any given $\ell$.  The price is a decrease in signal-to-noise.
Since the number of modes on the sky depends on $\ell$, such a strategy
would allow more accurate measurement of the large-scale power spectrum,
uncontaminated by the higher $k$ modes, if $\bar{n}_{\rm gal}$ was
sufficiently large.
Alternatively we could attempt to model the effect of baryon cooling.
The total theoretical uncertainty is then reduced to the uncertainty in
our modeling.  The treatment above is fairly crude but a more sophisticated
modeling, informed by numerical simulations, may be possible.

\section{Conclusions}

We have considered the effect of cooling of baryons on the small scale
mass power spectrum.  With a simple model for adiabatic compression we
can use the halo model to estimate the size of this effect on the 3D mass
power spectrum, and hence on the small angular scale convergence power
spectrum for weak lensing.
Our model contains numerous simplifying assumptions and provides at best
a ``factor of 2'' level calculation, but we find that baryons can alter the
mass profile non-negligibly on scales of $100\,h^{-1}$kpc which leads to
modifications of the angular power spectrum of $\kappa$ at
$\ell\simeq 3000-5000$ at the several percent level.
As indicated in Figure \ref{fig:dcl}, these modifications could be probed
by future weak lensing surveys and this would give us further insight
into the inner parts of dark matter halos.  It would also serve as an
extra source of theory uncertainty in constraining cosmological parameters
unless observational strategies to minimize it were implemented.
A refinement of this calculation, using modern numerical simulations
which model the heating and cooling effects of baryons, is clearly in order.

M.W. thanks Gary Bernstein, Dragan Huterer, Eric Linder, Chris Vale and
David Weinberg for useful conversations and comments on an earlier draft.
This research was supported by the NSF and NASA.

\end{document}